\documentclass[aps, amssymb, amsmath, reprint]{revtex4-1} 

\usepackage{graphicx}  
\usepackage{dcolumn}   
\usepackage{physics}
\usepackage{color}
\usepackage{appendix}
\usepackage{ulem}
\definecolor{orange}{rgb}{0.8, 0.3, 0}

\definecolor{blueviolet}{rgb}{0.2, 0.2, 0.6}
\usepackage[pdftex, 
bookmarks=true, 
colorlinks=true,
allcolors=blueviolet,
pdfstartview={FitH}, 
]{hyperref}

\newcommand{\gens}{g_\text{ens}}
\newcommand{\ktot}{\kappa_\text{tot}}

\begin{document}

\title{Scaling of self-stimulated spin echoes }

\author{S.~E.~de~Graaf$^{1}$}
\email{sdg@npl.co.uk}
\author{A.~Jayaraman$^{2}$}
\author{S.~E.~Kubatkin$^2$}
\author{A.~V.~Danilov$^2$}
\author{V.~Ranjan$^{1, 3}$}
\email{vranjan@tifrh.res.in}

\affiliation{$^1$National Physical Laboratory, Teddington TW11 0LW, United Kingdom }
\affiliation{$^2$Department of Microtechnology and Nanoscience MC2, Chalmers University of Technology, SE-41296 Goteborg, Sweden}
\affiliation{$^3$Present address: Tata Institute of Fundamental Research Hyderabad, Gopanpally, Hyderabad, 500046, Telangana, India}

\begin{abstract} 
Self-stimulated echoes have recently been reported in the high cooperativity and inhomogeneous coupling regime of spin ensembles with superconducting resonators. In this work, we study their relative amplitudes using echo-silencing made possible by a fast frequency tunable resonator. The highly anisotropic spin linewidth of Er$^{3+}$ electron spins in the CaWO$_4$ crystal also allows to study the dependence on spin-resonator ensemble cooperativity. It is demonstrated that self-stimulated echoes primarily result from a combination of two large control pulses and the echo preceding it.
\end{abstract}

\maketitle
In conventional magnetic resonance spectroscopy, stimulated echoes (STE) are known to occur when more than two control pulses are applied to spins. Stimulated echoes refocus the polarization grating stored on the longitudinal axis~\cite{schweiger_principles_2001}, in contrast to Hahn echoes which refocus the coherence generated on the transverse axes. In specific cases, Hahn echo emissions can themselves induce further evolution of spins and stimulate echo emissions. Although first observed in 1954~\cite{gordon_microwave_1958}, so called self-stimulated echoes (SSEs) have recently received renewed attention~\cite{debnath_self-stimulated_2020,weichselbaumer_echo_2020}.  This is because applications such as high sensitivity electron spin resonance spectroscopy~\cite{bienfait_reaching_2016, eichler_electron_2017, probst_inductive-detection_2017, ranjan_electron_2020, budoyo_electron_2020} and microwave quantum memories~\cite{julsgaard_quantum_2013, morton_storing_2018, grezes_multimode_2014, probst_microwave_2015, ranjan_multimode_2020, osullivan_random-access_2022} make use of 
 spin ensembles strongly coupled to superconducting resonators~\cite{imamoglu_cavity_2009, kubo_strong_2010, schuster_high-cooperativity_2010, abe_electron_2011, zhu_coherent_2011, amsuss_cavity_2011, ranjan_probing_2013, probst_anisotropic_2013, huebl_high_2013, sigillito_fast_2014, rose_coherent_2017, ball_loop-gap_2018}, a regime where SSEs are prevalent. 

The ensemble coupling of spins with a common resonator mode is quantified by the cooperativity $C=4g_\text{ens}^2 /\Gamma \ktot$, where $g_\text{ens} = g_0 \sqrt{N}$, $g_0$ the single spin-photon coupling strength, $\Gamma$ the inhomogeneous spin linewidth, $\ktot$ the total loss rate of the resonator and $N$ the number of spins. When $C \ll 1$, emitted echo fields are dissipated from the resonator before they could interact with spin ensemble again. On the other hand, when $C \gg 1$, a strong collective feedback effect of the emitted field on the spins, e.g. super-radiance~\cite{dicke_coherence_1954} and radiation damping~\cite{bloembergen_radiation_1954} can dominate the spin-dynamics. The intermediate regime of optimal impedance matching $C=1$ is especially relevant for maximum efficiency quantum memories \cite{afzelius_proposal_2013}. It is the purpose of this paper to experimentally study the scaling of self-stimulated echoes in these different regimes. Our study is, in particular, aided by the use a fast frequency tunable superconducting resonator \cite{mahashabde_fast_2020} for controlled emission of radiation into the resonator~\cite{ranjan_spin-echo_2022}.

\begin{figure}
    \centering
\includegraphics[width=\columnwidth]{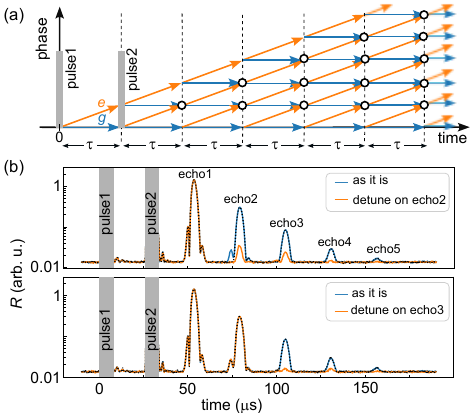} 
\caption{ \label{Fig1} Self-stimulated spin echoes (SSE). (a) A schematic of refocusing mechanism leading to self-stimulated echoes at $3\tau,~4\tau,~5\tau$... as originally described in Ref.~\onlinecite{debnath_self-stimulated_2020}. (b) Measured magnitude of echo trains using two pulses of same amplitude, duration $2~\mu$s and phase, and $\tau = 25~\mu$s. Two panels compare cases when the resonator is detuned to selectively suppress echo2 (top) or echo3 (bottom) emissions. Dashed curves in top (bottom) panels correspond to cases when the same resonator detuning pulse is applied between echo1 (echo2) and echo2 (echo3). Note that the dashed lines lie almost entirely on top of the solid lines, i.e. detuning in-between echoes has no effect. Measurements are done at $C = 3$. }
    \label{fig:my_label}
\end{figure}

Generation of SSEs can be understood using a simplified phase evolution in time, as proposed in Ref.~\onlinecite{debnath_self-stimulated_2020} and schematically presented in Fig.~\ref{Fig1}(a). When strong inhomogeneities of Rabi angles of spins exist, a control pulse brings the spins to different points on the Bloch sphere, which for simplicity can be decomposed into a subset of ground state ($g$) and excited state ($e$) amplitudes. A second control pulse at a time $\tau$ bifurcates the previous spin amplitudes into four subsets causing a refocusing at the time $2 \tau$ between two evolution trajectories, i.e. a conventional two-pulse Hahn echo. The emitted Hahn echo then itself acts like a pulse on spins such that new branches of spin evolution appear and additional refocusing events occur at a time $3\tau$. Subsequent echoes create more bifurcations and more refocusing events separated by $\tau$.  

We start our experimental studies by qualitatively verifying the sketch of Figure~\ref{Fig1}(a) which, in particular, illustrates that formation of SSEs requires phase evolution from all the pulses and echoes preceding it. Echo-trains measured using two control pulses of the same amplitude and phase are shown in Fig.~\ref{Fig1}(b). Note that the magnitude is plotted in the logarithmic scale. We observe that all subsequent echoes are suppressed when we detune the resonator frequency by an amount $\Delta\omega\gg \kappa$ to suppress the emission of echo2 (top panel)~\cite{ranjan_spin-echo_2022}. Applying the same duration detuning pulses between the echoes (dashed curves) produces no change thus proving that the detuning pulses do not generate significant phase noise to cause a suppression of echoes. Same observation of subsequent echo suppression is made when echo3 (bottom panel) is silenced. These suggest that contribution of 2-pulse refocusing to SSE, e.g. from pulse1 and echo1 in echo3, is small. In the following, we expand on the preceding observations and semi-quantitatively study the relative amplitudes of SSEs using in-situ control of radiation fields in the resonator and spin-resonator cooperativity. 

Our electron spins (with effective $S=1/2$) are provided by bulk doped Er$^{3+}$ substitutional ions in a CaWO$_4$ crystal with a nominal concentration of 50 ppm. The crystal is held with vacuum grease on a superconducting resonator of frequency $\omega_0/2\pi = 6.5$~GHz operating in the overcoupled regime with a loss rate of $\kappa_c/2\pi =1.9 \pm 0.1~$MHz. The bulk distribution of Er$^{3+}$ and narrow inductor width of $1~\mu$m naturally result in extremely inhomogeneous Rabi angles and benefit the formation of SSE. Two additional properties are relevant to this study. Firstly the kinetic inductance of the superconducting resonators (film thickness = 50~nm, inductor width = $0.5~\mu$m) made from NbN allows the resonance frequency to be rapidly tuned by passing a bias current through the inuctor strip of the resonator~\cite{mahashabde_fast_2020}. Secondly, it is possible to access different cooperativity $C$ in the same setup. This is because two isotopes of Er$^{3+}$, one without a nuclear spin $I=0$ ($77\%$) and the rest with $I=7/2$~\cite{rakonjac_long_2020} couple with different number of spins at different transitions. Moreover, fine tuning of $C$ is facilitated by the spin linewidth varying with the direction of the applied magnetic field as the highly anisotropic gyromagnetic tensor ($\gamma_{ab} = 117~\text{MHz/mT},~\gamma_{c} = 17~\text{MHz/mT}$)~\cite{enrique_optical_1971} responds to the charge-noise from crystal defects~\cite{mims_broadening_1966, dantec_twenty-threemillisecond_2021}. The experiments are performed at the base temperature of a dilution refrigerator at 20~mK, with the magnetic field aligned with the $c$-axis ($\phi \sim 0$) unless mentioned explicitly. More details of the experimental setup can be found in Ref.~\onlinecite{ranjan_spin-echo_2022}.

\begin{figure}[t!]
    \centering
\includegraphics[width=\columnwidth]{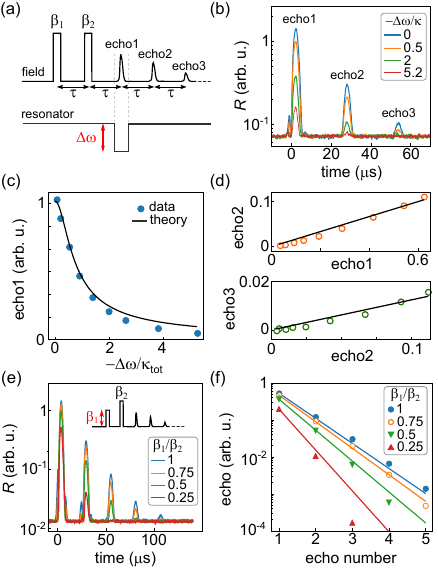} 
\caption{ \label{Fig2} SSE response versus intra-cavity field. (a) An experimental sequence consisting of two control pulses of flip angles $\beta_i$, and duration of $2~\mu$s with a $20~\mu$s long resonator detuning pulse across echo1 of varying $\Delta \omega$. (b) SSE traces at different $\Delta \omega$ and same flip angles $\beta= \beta_1 = \beta_2$. Larger noise floor is because of the larger measurement bandwidth $BW \sim 100~$MHz compared to other plots acquired at $BW$ of 2~MHz. (c) Measured (symbols) and theoretical (curve) echo amplitude against different resonator detunings. (d) Scaling of echo$\{2,3\}$ amplitudes (measured: symbols, fits: lines) with corresponding changes in echo\{1,2\}. (e) SSE traces for different flip angles $\beta_1$ of the first control pulse and fixed $\beta_2$. The sequence is shown in the inset. (f) SSE magnitude decay for different $\beta_1$ versus echo number. Solid lines are calculated from Eq.~\ref{eq:SSE}. For all plots $C = 3$.}
    \label{fig:my_label}
\end{figure}

The resonator tunability helps to control the back action of the echo field on the spins, that is to vary spin rotations during the echo emission, and study the amplitudes of subsequent SSE. As shown in the sketch of Fig.~\ref{Fig2}(a), two pulses of the same amplitude and phase are applied and the resonator detuned for $20~\mu$s, a time longer than the echo duration, with varying $\Delta \omega$ around echo1. Figure~\ref{Fig2}(b) shows the corresponding echo train traces (acquired at a large demodulation bandwidth of 100~MHz to account for the relatively large total loss rate $\ktot \sim 7~$MHz) near the $I=0$ transition with $C = 3$ (see further below). The variation of echo1 magnitude versus normalized resonator detuning $-\Delta \omega/ \ktot$ is plotted in Fig.~\ref{Fig2}(c) and the observed decay is well accounted for by the resonator filtering function $(\ktot/2)/\sqrt{\Delta \omega^2+\ktot^2/4}$~\cite{ranjan_spin-echo_2022}. Similar to Fig.~\ref{Fig1}(b), subsequent echoes, echo2 and echo3 are progressively suppressed. To quantify their relative suppression, we plot the amplitude of echo2 and echo3 as a function of echo1  and echo2, respectively, in Fig.~\ref{Fig2}(d). A linear dependence (proportionality constant 0.16 and 0.12, respectively) describes the echo2 and echo3 data well. 

Full quantitative understanding  of the scaling of SSE is challenging due to the lack of knowledge of exact spin frequency detuning and coupling strength distribution. Here, we use a minimalist model to explain the scaling of echo2 and echo3 using classical Bloch theory. Three pulses with arbitrary flip angles $\beta_i$ produce a STE with an amplitude proportional to $\sin(\beta_1)  \sin(\beta_2)  \sin(\beta_3)$~\cite{schweiger_principles_2001}, where we assume pulse delay $\tau \ll T_2, ~T_1$. Using control pulses of the same Rabi angle $\beta$ and the fact that resulting echo1 fields are relatively much smaller, the resulting spin rotations from back action is $\text{sin}(\theta_1) \approx \theta_1$. Then the STE contribution of echo2 is equal to $\theta_1 \sin^2(\beta)$, where $\theta_i$ denotes the much smaller rotation angle from echo back action. Similarly, the 2-pulse Hahn echo contribution of echo2 (from pulse2 and echo1) is proportional to $\theta_1^2 \sin(\beta)$. The latter is smaller in magnitude than the STE contribution as long as $\beta \gg \theta_1$. Thus linear scaling of echo2 with echo1 can be established. Similar arguments can be made for echo3 to show that the dominating contribution comes from a 3-pulse STE from pulse1, pulse2 and echo2, with a resulting echo3 proportional to $\theta_2 \sin^2(\beta)$. The proportionality constant extracted from slopes in two cases is found to be similar, 0.16 and 0.12, as expected from the model.

Overall our observations in Fig.~\ref{Fig2}(d) suggest that a SSE primarily consists of a 3-pulse STE from two large control pulses and the weak echo field preceding it. Barring common prefactors, we can thus quantify the magnitude of the $(i+1)^\text{th}$ echo in the limit of $\tau \ll T_1$ as  

\begin{equation} \label{eq:SSE}
    A_\text{echo}^{i+1} \equiv \eta A_\text{echo}^i \text{sin} (\beta_1) \text{sin}(\beta_2),
\end{equation}
where $i>0$ is a positive integer, and a scaling factor $\eta^2$ captures the fraction of power transferred to spins during the formation of an echo. To verify this equation further, we acquired SSE traces by varying the flip angle of the first control pulse $\beta_1$ (Fig.~\ref{Fig2}(e)), while keeping $\beta_2$ fixed. Their decay is plotted in Fig.~\ref{Fig2}(f). It has been previously shown that for strongly inhomogeneous Rabi angles in spin systems coupled to small mode volume resonators, spins for which pulse amplitudes amount to $\pi/2$ and $\pi$ contribute maximum to the Hahn echo~\cite{ranjan_pulsed_2020, osullivan_spin-resonance_2020, dantec_twenty-threemillisecond_2021}. This allows us to set $\beta_2 = 90^\circ$, and proportionally vary $\beta_1$ using the ratio of pulse amplitudes $\beta_2/\beta_1$. The SSE decays calculated from Eq.~\ref{eq:SSE} are plotted as solid lines in Fig.~\ref{Fig2}(f), and show an excellent agreement with measurements using the same scaling parameter $\eta=0.21$ across the entire dataset. Moreover, $\eta \sin^2(\beta) \approx 0.21$ is close to the measured slope in Fig.~\ref{Fig2}(d) acquired under the same experimental conditions. 

\begin{figure}
    \centering
\includegraphics[width=\columnwidth]{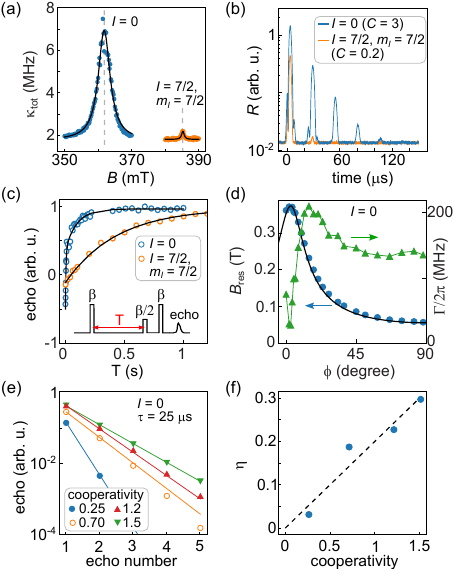}
\caption{ \label{Fig3} SSE response versus spin-resonator cooperativity. (a) Continuous wave spectroscopy near two Er$^{3+}$ transitions $I=0$ and $I=7/2,~m_I=7/2$ at zero angle (measured: symbols, fit:lines). (b) Spin energy relaxation (measured: symbols, fit:lines) using inversion recovery sequences for the two transitions at zero angle. (c) echo response using the control pulses with same power. (d) Spin resonance position (left axis, measured: symbols, theory: line) and spin linewidth (right axis) for the $I=0$ transition extracted from continuous wave spectroscopy. The magnetic field angle $\phi$ is relative to the $c$-axis of CaWO$_4$. (e) Decay of SSE magnitudes for different cooperativity, but similar $\kappa_\text{tot}$, obtained at different $\phi$ for the $I=0$ transition. Solid lines are calculated using Eq.~\ref{eq:SSE}. (f) scaling of the extracted scaling parameter $\eta$ as a function of $C$. Dashed line is a guide to the eye.}
    \label{fig:my_label}
\end{figure}

We now study the dependence of SSE amplitudes on spin ensemble-resonator cooperativity. To this end, we identify two transitions in the spectrum [Fig.~\ref{Fig3}(a)] belonging to nuclear spin isotopes $I=0$ and $I=7/2$ ($m_I=7/2$ is the nuclear spin projection on the magnetic field axis). From fits performed to $\ktot = \kappa + \gens^2 \Gamma/(\Delta \omega_s^2+ \Gamma^2/4)  $~\cite{schuster_high-cooperativity_2010}, we find coupling strengths $\gens/2\pi = 10 \pm 1$~MHz and $1.2 \pm 0.1$~MHz, and spin linewidths $\Gamma/2\pi = 76\pm 5$~MHz and $15\pm 1$~MHz, and corresponding cooperativity $C = $ 3 and 0.2, respectively. Here $\Delta \omega_s$ is the magnetic field dependent detuning of the spin transition frequency from the resonator. The difference in number of spins is consistent with the isotope and seven sub-level ground state populations in the $I=7/2$ manifold. Echo response measured using the same control pulses ($2~\mu$s in duration, such that pulse bandwidth $\ll \Gamma$) at two transitions [Fig.~\ref{Fig3}(b)] shows strongly suppressed or absent SSE for the case of $C \ll 1$, and supports similar observations made in Ref.~\onlinecite{weichselbaumer_echo_2020}. 

To investigate differences of spin dynamics between two Er isotopes, the spin-relaxation time is measured using an inversion recovery sequence (Fig.~\ref{Fig3}(c)). For $I=7/2$, we observe an exponential recovery with a decay constant $T_1 = 440 \pm 11~$ms, a value consistent with a direct phonon process~\cite{ranjan_spin-echo_2022, dantec_notitle_2022}. In contrast, we observe a bi-exponential recovery for $I=0$, with decay constants $T_1^\text{fast} = 4.7 \pm 0.6~$ms and $T_1^\text{slow} = 97 \pm 12~$ms. However, neither of the two values is compatible with a direct-phonon process (scaling as $1/B^5$), suggesting that a combination of strong collective radiative effects, i.e. superradiance and spatial spin diffusion across the low mode volume resonator~\cite{ranjan_probing_2013} could be responsible. The role of incoherent radiation from enhanced spin relaxation towards formation of SSEs can, however, be ruled out as resonator detuning pulses of duration $20~\mu$s applied in-between the echoes (dashed curves in Fig.~\ref{Fig1}(b)) do not alter the subsequent echoes. We also measure spin coherence times $T_2$ at two transitions and find the contrasting SSE amplitudes to not be related to the relative $T_2$ times. In fact, $T_2=2.5~$ms for $I=7/2$ is four times longer compared to that for $I=0$ and possibly limited by instantaneous diffusion~\cite{rancic_electron-spin_2022-1, alexander_coherent_2022}. 

Another control of cooperativity is achieved by different $\Gamma$ of the spin ensemble obtained when rotating the applied magnetic field with respect to the $c$-axis of the crystal. Figure~\ref{Fig3}(d) shows measured magnetic field $B_\text{res}$ at which the $I=0$ transition is resonant with the resonator (left axis) and the extracted spin linewidth $\Gamma$ (right axis). The $B_\text{res}$ positions agree with the spin Hamiltonian of Er$^{3+}$ with a reasonable misalignment angle of $2.5^\circ$ from the true $c$-axis. We observed a change in $\Gamma$ from 30~MHz at $\phi = 2.5^\circ$ to 210~MHz at $\phi = 21^\circ$. Similar observations have been made previously~\cite{mims_broadening_1966, dantec_twenty-threemillisecond_2021} and attributed to a combination of local electric fields from charge defects, charge compensation and lack of inversion symmetry at the substitutional Ca$^{2+}$ sites. On the other hand, the extracted ensemble coupling strength $\gens$ decreases by only $10\%$ in this $\phi$ range. The small variation in $\gens$ is consistent with $g_0$ calculated from the anisotropic gyromagnetic tensor, and the fact that the same number of spins are coupled to the resonator due to the bulk distribution of Er$^{3+}$.

For SSE measurements, we choose slightly off-resonant $B$ fields at different $\phi$ to achieve a maximum echo amplitude~\cite{alexander_coherent_2022} and somewhat similar $\ktot$ values (between 2.8~MHz and 4~MHz) for better comparison. The SSE magnitudes measured with the same control pulses and delay $\tau = 25~\mu$s are plotted as a function of echo number in Fig.~\ref{Fig3}(e) for different cooperativity $C$. We note that the off-resonant $C$ is extracted by comparing the intra-cavity field measured at a repetition rate $\gamma_\text{rep} \ll T_1$ (spins saturated) with that taken at $\gamma_\text{rep} \gg T_1$ (spins polarized)~\cite{grezes_multimode_2014, ranjan_multimode_2020}. For all values of $C$, we observe an exponential decay of echo amplitudes, similar to Fig.~\ref{Fig2}(e) and Ref.~\onlinecite{weichselbaumer_echo_2020}. For extracting $\eta$ using Eq.~\ref{eq:SSE}, once again we set $\beta_{1,2} = 90^\circ$ to select the spins that maximally contribute to SSE amplitudes. The calculated SSE decays and corresponding $\eta$ for different $C$ are plotted in Fig.~\ref{Fig3}(e, f). Interestingly, the scaling parameter $\eta$ increases with $C$ in an apparent linear fashion. In contrast, $\eta=0.21$ extracted in Fig.~\ref{Fig2}(f) at a larger $C=3$ is smaller than $\eta=0.3$ for $C=1.5$ in Fig.~\ref{Fig3}(e), suggesting the role of larger $\ktot$ in the smaller spin rotation during echoes, crudeness of the model and a more complex dependence of $\eta$ on $C$. 

In conclusion, we have used control of intra-cavity field, in particular through echo-silencing, and cooperativity tuning to study scaling of self-stimulated echoes in a strongly inhomogeneously coupled spin ensemble to a small mode volume superconducting resonator. Our results demonstrate that the amplitude of a self-stimulated echo primarily arises from a three pulse stimulated echo using two large control pulses and the preceding echo field. Further studies will target a larger range of $C$, especially at a fixed $\ktot$, to map out the scaling and decay of SSE amplitudes against $C$. STE and SSE in combination with phase imprinting~\cite{wu_storage_2010, ranjan_spin-echo_2022} could also be used to implement selective in situ magnetic resonance techniques such as diffusion spectroscopy and imaging \cite{burstein_stimulated_1996}.

We acknowledge the support from the UK Department for Science, Innovation and Technology through the UK national quantum technologies program. S.D.G. acknowledges support by the Engineering and Physical Sciences Research Council (EPSRC) (Grant Number EP/W027526/1). The Chalmers group acknowledges the support from the Swedish Research Council (VR) (Grant Agreements No. 2019-05480 and No. 2020-04393), EU H2020 European Microkelvin Platform (Grant Agreement No. 824109), and from Knut and Alice Wallenberg Foundation via the Wallenberg centre for Quantum Technology (WACQT). This work was performed in part at Myfab Chalmers.
\bibliography{QuantumMemory}

\begin{thebibliography}{43}%
\makeatletter
\providecommand \@ifxundefined [1]{%
 \@ifx{#1\undefined}
}%
\providecommand \@ifnum [1]{%
 \ifnum #1\expandafter \@firstoftwo
 \else \expandafter \@secondoftwo
 \fi
}%
\providecommand \@ifx [1]{%
 \ifx #1\expandafter \@firstoftwo
 \else \expandafter \@secondoftwo
 \fi
}%
\providecommand \natexlab [1]{#1}%
\providecommand \enquote  [1]{``#1''}%
\providecommand \bibnamefont  [1]{#1}%
\providecommand \bibfnamefont [1]{#1}%
\providecommand \citenamefont [1]{#1}%
\providecommand \href@noop [0]{\@secondoftwo}%
\providecommand \href [0]{\begingroup \@sanitize@url \@href}%
\providecommand \@href[1]{\@@startlink{#1}\@@href}%
\providecommand \@@href[1]{\endgroup#1\@@endlink}%
\providecommand \@sanitize@url [0]{\catcode `\\12\catcode `\$12\catcode
  `\&12\catcode `\#12\catcode `\^12\catcode `\_12\catcode `\%12\relax}%
\providecommand \@@startlink[1]{}%
\providecommand \@@endlink[0]{}%
\providecommand \url  [0]{\begingroup\@sanitize@url \@url }%
\providecommand \@url [1]{\endgroup\@href {#1}{\urlprefix }}%
\providecommand \urlprefix  [0]{URL }%
\providecommand \Eprint [0]{\href }%
\providecommand \doibase [0]{http://dx.doi.org/}%
\providecommand \selectlanguage [0]{\@gobble}%
\providecommand \bibinfo  [0]{\@secondoftwo}%
\providecommand \bibfield  [0]{\@secondoftwo}%
\providecommand \translation [1]{[#1]}%
\providecommand \BibitemOpen [0]{}%
\providecommand \bibitemStop [0]{}%
\providecommand \bibitemNoStop [0]{.\EOS\space}%
\providecommand \EOS [0]{\spacefactor3000\relax}%
\providecommand \BibitemShut  [1]{\csname bibitem#1\endcsname}%
\let\auto@bib@innerbib\@empty
\bibitem [{\citenamefont {Schweiger}\ and\ \citenamefont
  {Jeschke}(2001)}]{schweiger_principles_2001}%
  \BibitemOpen
  \bibfield  {author} {\bibinfo {author} {\bibfnamefont {A.}~\bibnamefont
  {Schweiger}}\ and\ \bibinfo {author} {\bibfnamefont {G.}~\bibnamefont
  {Jeschke}},\ }\href@noop {} {\emph {\bibinfo {title} {Principles of pulse
  electron paramagnetic resonance}}}\ (\bibinfo  {publisher} {Oxford University
  Press},\ \bibinfo {year} {2001})\BibitemShut {NoStop}%
\bibitem [{\citenamefont {Gordon}\ and\ \citenamefont
  {Bowers}(1958)}]{gordon_microwave_1958}%
  \BibitemOpen
  \bibfield  {author} {\bibinfo {author} {\bibfnamefont {J.~P.}\ \bibnamefont
  {Gordon}}\ and\ \bibinfo {author} {\bibfnamefont {K.~D.}\ \bibnamefont
  {Bowers}},\ }\href {\doibase 10.1103/PhysRevLett.1.368} {\bibfield  {journal}
  {\bibinfo  {journal} {Physical Review Letters}\ }\textbf {\bibinfo {volume}
  {1}},\ \bibinfo {pages} {368} (\bibinfo {year} {1958})}\BibitemShut {NoStop}%
\bibitem [{\citenamefont {Debnath}\ \emph {et~al.}(2020)\citenamefont
  {Debnath}, \citenamefont {Dold}, \citenamefont {Morton},\ and\ \citenamefont
  {M{\o}lmer}}]{debnath_self-stimulated_2020}%
  \BibitemOpen
  \bibfield  {author} {\bibinfo {author} {\bibfnamefont {K.}~\bibnamefont
  {Debnath}}, \bibinfo {author} {\bibfnamefont {G.}~\bibnamefont {Dold}},
  \bibinfo {author} {\bibfnamefont {J.~J.~L.}\ \bibnamefont {Morton}}, \ and\
  \bibinfo {author} {\bibfnamefont {K.}~\bibnamefont {M{\o}lmer}},\ }\href
  {\doibase 10.1103/PhysRevLett.125.137702} {\bibfield  {journal} {\bibinfo
  {journal} {Physical Review Letters}\ }\textbf {\bibinfo {volume} {125}},\
  \bibinfo {pages} {137702} (\bibinfo {year} {2020})}\BibitemShut {NoStop}%
\bibitem [{\citenamefont {Weichselbaumer}\ \emph {et~al.}(2020)\citenamefont
  {Weichselbaumer}, \citenamefont {Zens}, \citenamefont {Zollitsch},
  \citenamefont {Brandt}, \citenamefont {Rotter}, \citenamefont {Gross},\ and\
  \citenamefont {Huebl}}]{weichselbaumer_echo_2020}%
  \BibitemOpen
  \bibfield  {author} {\bibinfo {author} {\bibfnamefont {S.}~\bibnamefont
  {Weichselbaumer}}, \bibinfo {author} {\bibfnamefont {M.}~\bibnamefont
  {Zens}}, \bibinfo {author} {\bibfnamefont {C.~W.}\ \bibnamefont {Zollitsch}},
  \bibinfo {author} {\bibfnamefont {M.~S.}\ \bibnamefont {Brandt}}, \bibinfo
  {author} {\bibfnamefont {S.}~\bibnamefont {Rotter}}, \bibinfo {author}
  {\bibfnamefont {R.}~\bibnamefont {Gross}}, \ and\ \bibinfo {author}
  {\bibfnamefont {H.}~\bibnamefont {Huebl}},\ }\href {\doibase
  10.1103/PhysRevLett.125.137701} {\bibfield  {journal} {\bibinfo  {journal}
  {Physical Review Letters}\ }\textbf {\bibinfo {volume} {125}},\ \bibinfo
  {pages} {137701} (\bibinfo {year} {2020})}\BibitemShut {NoStop}%
\bibitem [{\citenamefont {Bienfait}\ \emph {et~al.}(2016)\citenamefont
  {Bienfait}, \citenamefont {Pla}, \citenamefont {Kubo}, \citenamefont {Stern},
  \citenamefont {Zhou}, \citenamefont {Lo}, \citenamefont {Weis}, \citenamefont
  {Schenkel}, \citenamefont {Thewalt}, \citenamefont {Vion}, \citenamefont
  {Esteve}, \citenamefont {Julsgaard}, \citenamefont {M{\o}lmer}, \citenamefont
  {Morton},\ and\ \citenamefont {Bertet}}]{bienfait_reaching_2016}%
  \BibitemOpen
  \bibfield  {author} {\bibinfo {author} {\bibfnamefont {A.}~\bibnamefont
  {Bienfait}}, \bibinfo {author} {\bibfnamefont {J.~J.}\ \bibnamefont {Pla}},
  \bibinfo {author} {\bibfnamefont {Y.}~\bibnamefont {Kubo}}, \bibinfo {author}
  {\bibfnamefont {M.}~\bibnamefont {Stern}}, \bibinfo {author} {\bibfnamefont
  {X.}~\bibnamefont {Zhou}}, \bibinfo {author} {\bibfnamefont {C.~C.}\
  \bibnamefont {Lo}}, \bibinfo {author} {\bibfnamefont {C.~D.}\ \bibnamefont
  {Weis}}, \bibinfo {author} {\bibfnamefont {T.}~\bibnamefont {Schenkel}},
  \bibinfo {author} {\bibfnamefont {M.~L.~W.}\ \bibnamefont {Thewalt}},
  \bibinfo {author} {\bibfnamefont {D.}~\bibnamefont {Vion}}, \bibinfo {author}
  {\bibfnamefont {D.}~\bibnamefont {Esteve}}, \bibinfo {author} {\bibfnamefont
  {B.}~\bibnamefont {Julsgaard}}, \bibinfo {author} {\bibfnamefont
  {K.}~\bibnamefont {M{\o}lmer}}, \bibinfo {author} {\bibfnamefont {J.~J.~L.}\
  \bibnamefont {Morton}}, \ and\ \bibinfo {author} {\bibfnamefont
  {P.}~\bibnamefont {Bertet}},\ }\href {\doibase 10.1038/nnano.2015.282}
  {\bibfield  {journal} {\bibinfo  {journal} {Nature Nanotechnology}\ }\textbf
  {\bibinfo {volume} {11}},\ \bibinfo {pages} {253} (\bibinfo {year}
  {2016})}\BibitemShut {NoStop}%
\bibitem [{\citenamefont {Eichler}\ \emph {et~al.}(2017)\citenamefont
  {Eichler}, \citenamefont {Sigillito}, \citenamefont {Lyon},\ and\
  \citenamefont {Petta}}]{eichler_electron_2017}%
  \BibitemOpen
  \bibfield  {author} {\bibinfo {author} {\bibfnamefont {C.}~\bibnamefont
  {Eichler}}, \bibinfo {author} {\bibfnamefont {A.~J.}\ \bibnamefont
  {Sigillito}}, \bibinfo {author} {\bibfnamefont {S.~A.}\ \bibnamefont {Lyon}},
  \ and\ \bibinfo {author} {\bibfnamefont {J.~R.}\ \bibnamefont {Petta}},\
  }\href {\doibase 10.1103/PhysRevLett.118.037701} {\bibfield  {journal}
  {\bibinfo  {journal} {Physical Review Letters}\ }\textbf {\bibinfo {volume}
  {118}},\ \bibinfo {pages} {037701} (\bibinfo {year} {2017})}\BibitemShut
  {NoStop}%
\bibitem [{\citenamefont {Probst}\ \emph {et~al.}(2017)\citenamefont {Probst},
  \citenamefont {Bienfait}, \citenamefont {Campagne-Ibarcq}, \citenamefont
  {Pla}, \citenamefont {Albanese}, \citenamefont {Da~Silva~Barbosa},
  \citenamefont {Schenkel}, \citenamefont {Vion}, \citenamefont {Esteve},
  \citenamefont {M{\o}lmer}, \citenamefont {Morton}, \citenamefont {Heeres},\
  and\ \citenamefont {Bertet}}]{probst_inductive-detection_2017}%
  \BibitemOpen
  \bibfield  {author} {\bibinfo {author} {\bibfnamefont {S.}~\bibnamefont
  {Probst}}, \bibinfo {author} {\bibfnamefont {A.}~\bibnamefont {Bienfait}},
  \bibinfo {author} {\bibfnamefont {P.}~\bibnamefont {Campagne-Ibarcq}},
  \bibinfo {author} {\bibfnamefont {J.~J.}\ \bibnamefont {Pla}}, \bibinfo
  {author} {\bibfnamefont {B.}~\bibnamefont {Albanese}}, \bibinfo {author}
  {\bibfnamefont {J.~F.}\ \bibnamefont {Da~Silva~Barbosa}}, \bibinfo {author}
  {\bibfnamefont {T.}~\bibnamefont {Schenkel}}, \bibinfo {author}
  {\bibfnamefont {D.}~\bibnamefont {Vion}}, \bibinfo {author} {\bibfnamefont
  {D.}~\bibnamefont {Esteve}}, \bibinfo {author} {\bibfnamefont
  {K.}~\bibnamefont {M{\o}lmer}}, \bibinfo {author} {\bibfnamefont {J.~J.~L.}\
  \bibnamefont {Morton}}, \bibinfo {author} {\bibfnamefont {R.}~\bibnamefont
  {Heeres}}, \ and\ \bibinfo {author} {\bibfnamefont {P.}~\bibnamefont
  {Bertet}},\ }\href {\doibase 10.1063/1.5002540} {\bibfield  {journal}
  {\bibinfo  {journal} {Applied Physics Letters}\ }\textbf {\bibinfo {volume}
  {111}},\ \bibinfo {pages} {202604} (\bibinfo {year} {2017})}\BibitemShut
  {NoStop}%
\bibitem [{\citenamefont {Ranjan}\ \emph
  {et~al.}(2020{\natexlab{a}})\citenamefont {Ranjan}, \citenamefont {Probst},
  \citenamefont {Albanese}, \citenamefont {Schenkel}, \citenamefont {Vion},
  \citenamefont {Esteve}, \citenamefont {Morton},\ and\ \citenamefont
  {Bertet}}]{ranjan_electron_2020}%
  \BibitemOpen
  \bibfield  {author} {\bibinfo {author} {\bibfnamefont {V.}~\bibnamefont
  {Ranjan}}, \bibinfo {author} {\bibfnamefont {S.}~\bibnamefont {Probst}},
  \bibinfo {author} {\bibfnamefont {B.}~\bibnamefont {Albanese}}, \bibinfo
  {author} {\bibfnamefont {T.}~\bibnamefont {Schenkel}}, \bibinfo {author}
  {\bibfnamefont {D.}~\bibnamefont {Vion}}, \bibinfo {author} {\bibfnamefont
  {D.}~\bibnamefont {Esteve}}, \bibinfo {author} {\bibfnamefont {J.~J.~L.}\
  \bibnamefont {Morton}}, \ and\ \bibinfo {author} {\bibfnamefont
  {P.}~\bibnamefont {Bertet}},\ }\href {\doibase 10.1063/5.0004322} {\bibfield
  {journal} {\bibinfo  {journal} {Applied Physics Letters}\ }\textbf {\bibinfo
  {volume} {116}},\ \bibinfo {pages} {184002} (\bibinfo {year}
  {2020}{\natexlab{a}})}\BibitemShut {NoStop}%
\bibitem [{\citenamefont {Budoyo}\ \emph {et~al.}(2020)\citenamefont {Budoyo},
  \citenamefont {Kakuyanagi}, \citenamefont {Toida}, \citenamefont
  {Matsuzaki},\ and\ \citenamefont {Saito}}]{budoyo_electron_2020}%
  \BibitemOpen
  \bibfield  {author} {\bibinfo {author} {\bibfnamefont {R.~P.}\ \bibnamefont
  {Budoyo}}, \bibinfo {author} {\bibfnamefont {K.}~\bibnamefont {Kakuyanagi}},
  \bibinfo {author} {\bibfnamefont {H.}~\bibnamefont {Toida}}, \bibinfo
  {author} {\bibfnamefont {Y.}~\bibnamefont {Matsuzaki}}, \ and\ \bibinfo
  {author} {\bibfnamefont {S.}~\bibnamefont {Saito}},\ }\href {\doibase
  10.1063/1.5144722} {\bibfield  {journal} {\bibinfo  {journal} {Applied
  Physics Letters}\ }\textbf {\bibinfo {volume} {116}},\ \bibinfo {pages}
  {194001} (\bibinfo {year} {2020})}\BibitemShut {NoStop}%
\bibitem [{\citenamefont {Julsgaard}\ \emph {et~al.}(2013)\citenamefont
  {Julsgaard}, \citenamefont {Grezes}, \citenamefont {Bertet},\ and\
  \citenamefont {Molmer}}]{julsgaard_quantum_2013}%
  \BibitemOpen
  \bibfield  {author} {\bibinfo {author} {\bibfnamefont {B.}~\bibnamefont
  {Julsgaard}}, \bibinfo {author} {\bibfnamefont {C.}~\bibnamefont {Grezes}},
  \bibinfo {author} {\bibfnamefont {P.}~\bibnamefont {Bertet}}, \ and\ \bibinfo
  {author} {\bibfnamefont {K.}~\bibnamefont {Molmer}},\ }\href {\doibase
  10.1103/PhysRevLett.110.250503} {\bibfield  {journal} {\bibinfo  {journal}
  {Physical Review Letters}\ }\textbf {\bibinfo {volume} {110}} (\bibinfo
  {year} {2013}),\ 10.1103/PhysRevLett.110.250503}\BibitemShut {NoStop}%
\bibitem [{\citenamefont {Morton}\ and\ \citenamefont
  {Bertet}(2018)}]{morton_storing_2018}%
  \BibitemOpen
  \bibfield  {author} {\bibinfo {author} {\bibfnamefont {J.~J.~L.}\
  \bibnamefont {Morton}}\ and\ \bibinfo {author} {\bibfnamefont
  {P.}~\bibnamefont {Bertet}},\ }\href {\doibase 10.1016/j.jmr.2017.11.015}
  {\bibfield  {journal} {\bibinfo  {journal} {Journal of Magnetic Resonance}\
  }\textbf {\bibinfo {volume} {287}},\ \bibinfo {pages} {128} (\bibinfo {year}
  {2018})}\BibitemShut {NoStop}%
\bibitem [{\citenamefont {Grezes}\ \emph {et~al.}(2014)\citenamefont {Grezes},
  \citenamefont {Julsgaard}, \citenamefont {Kubo}, \citenamefont {Stern},
  \citenamefont {Umeda}, \citenamefont {Isoya}, \citenamefont {Sumiya},
  \citenamefont {Abe}, \citenamefont {Onoda}, \citenamefont {Ohshima},
  \citenamefont {Jacques}, \citenamefont {Esteve}, \citenamefont {Vion},
  \citenamefont {Esteve}, \citenamefont {Molmer},\ and\ \citenamefont
  {Bertet}}]{grezes_multimode_2014}%
  \BibitemOpen
  \bibfield  {author} {\bibinfo {author} {\bibfnamefont {C.}~\bibnamefont
  {Grezes}}, \bibinfo {author} {\bibfnamefont {B.}~\bibnamefont {Julsgaard}},
  \bibinfo {author} {\bibfnamefont {Y.}~\bibnamefont {Kubo}}, \bibinfo {author}
  {\bibfnamefont {M.}~\bibnamefont {Stern}}, \bibinfo {author} {\bibfnamefont
  {T.}~\bibnamefont {Umeda}}, \bibinfo {author} {\bibfnamefont
  {J.}~\bibnamefont {Isoya}}, \bibinfo {author} {\bibfnamefont
  {H.}~\bibnamefont {Sumiya}}, \bibinfo {author} {\bibfnamefont
  {H.}~\bibnamefont {Abe}}, \bibinfo {author} {\bibfnamefont {S.}~\bibnamefont
  {Onoda}}, \bibinfo {author} {\bibfnamefont {T.}~\bibnamefont {Ohshima}},
  \bibinfo {author} {\bibfnamefont {V.}~\bibnamefont {Jacques}}, \bibinfo
  {author} {\bibfnamefont {J.}~\bibnamefont {Esteve}}, \bibinfo {author}
  {\bibfnamefont {D.}~\bibnamefont {Vion}}, \bibinfo {author} {\bibfnamefont
  {D.}~\bibnamefont {Esteve}}, \bibinfo {author} {\bibfnamefont
  {K.}~\bibnamefont {Molmer}}, \ and\ \bibinfo {author} {\bibfnamefont
  {P.}~\bibnamefont {Bertet}},\ }\href {\doibase 10.1103/PhysRevX.4.021049}
  {\bibfield  {journal} {\bibinfo  {journal} {Physical Review X}\ }\textbf
  {\bibinfo {volume} {4}} (\bibinfo {year} {2014}),\
  10.1103/PhysRevX.4.021049}\BibitemShut {NoStop}%
\bibitem [{\citenamefont {Probst}\ \emph {et~al.}(2015)\citenamefont {Probst},
  \citenamefont {Rotzinger}, \citenamefont {Ustinov},\ and\ \citenamefont
  {Bushev}}]{probst_microwave_2015}%
  \BibitemOpen
  \bibfield  {author} {\bibinfo {author} {\bibfnamefont {S.}~\bibnamefont
  {Probst}}, \bibinfo {author} {\bibfnamefont {H.}~\bibnamefont {Rotzinger}},
  \bibinfo {author} {\bibfnamefont {A.~V.}\ \bibnamefont {Ustinov}}, \ and\
  \bibinfo {author} {\bibfnamefont {P.~A.}\ \bibnamefont {Bushev}},\ }\href
  {\doibase 10.1103/PhysRevB.92.014421} {\bibfield  {journal} {\bibinfo
  {journal} {Physical Review B}\ }\textbf {\bibinfo {volume} {92}},\ \bibinfo
  {pages} {014421} (\bibinfo {year} {2015})}\BibitemShut {NoStop}%
\bibitem [{\citenamefont {Ranjan}\ \emph
  {et~al.}(2020{\natexlab{b}})\citenamefont {Ranjan}, \citenamefont
  {O{\textquoteright}Sullivan}, \citenamefont {Albertinale}, \citenamefont
  {Albanese}, \citenamefont {Chaneli{\`e}re}, \citenamefont {Schenkel},
  \citenamefont {Vion}, \citenamefont {Esteve}, \citenamefont {Flurin},
  \citenamefont {Morton},\ and\ \citenamefont
  {Bertet}}]{ranjan_multimode_2020}%
  \BibitemOpen
  \bibfield  {author} {\bibinfo {author} {\bibfnamefont {V.}~\bibnamefont
  {Ranjan}}, \bibinfo {author} {\bibfnamefont {J.}~\bibnamefont
  {O{\textquoteright}Sullivan}}, \bibinfo {author} {\bibfnamefont
  {E.}~\bibnamefont {Albertinale}}, \bibinfo {author} {\bibfnamefont
  {B.}~\bibnamefont {Albanese}}, \bibinfo {author} {\bibfnamefont
  {T.}~\bibnamefont {Chaneli{\`e}re}}, \bibinfo {author} {\bibfnamefont
  {T.}~\bibnamefont {Schenkel}}, \bibinfo {author} {\bibfnamefont
  {D.}~\bibnamefont {Vion}}, \bibinfo {author} {\bibfnamefont {D.}~\bibnamefont
  {Esteve}}, \bibinfo {author} {\bibfnamefont {E.}~\bibnamefont {Flurin}},
  \bibinfo {author} {\bibfnamefont {J.~J.~L.}\ \bibnamefont {Morton}}, \ and\
  \bibinfo {author} {\bibfnamefont {P.}~\bibnamefont {Bertet}},\ }\href
  {\doibase 10.1103/PhysRevLett.125.210505} {\bibfield  {journal} {\bibinfo
  {journal} {Physical Review Letters}\ }\textbf {\bibinfo {volume} {125}},\
  \bibinfo {pages} {210505} (\bibinfo {year} {2020}{\natexlab{b}})}\BibitemShut
  {NoStop}%
\bibitem [{\citenamefont {O{\textquoteright}Sullivan}\ \emph
  {et~al.}(2022)\citenamefont {O{\textquoteright}Sullivan}, \citenamefont
  {Kennedy}, \citenamefont {Debnath}, \citenamefont {Alexander}, \citenamefont
  {Zollitsch}, \citenamefont {{\v S}im{\.e}nas}, \citenamefont {Hashim},
  \citenamefont {Thomas}, \citenamefont {Withington}, \citenamefont {Siddiqi},
  \citenamefont {M{\o}lmer},\ and\ \citenamefont
  {Morton}}]{osullivan_random-access_2022}%
  \BibitemOpen
  \bibfield  {author} {\bibinfo {author} {\bibfnamefont {J.}~\bibnamefont
  {O{\textquoteright}Sullivan}}, \bibinfo {author} {\bibfnamefont {O.~W.}\
  \bibnamefont {Kennedy}}, \bibinfo {author} {\bibfnamefont {K.}~\bibnamefont
  {Debnath}}, \bibinfo {author} {\bibfnamefont {J.}~\bibnamefont {Alexander}},
  \bibinfo {author} {\bibfnamefont {C.~W.}\ \bibnamefont {Zollitsch}}, \bibinfo
  {author} {\bibfnamefont {M.}~\bibnamefont {{\v S}im{\.e}nas}}, \bibinfo
  {author} {\bibfnamefont {A.}~\bibnamefont {Hashim}}, \bibinfo {author}
  {\bibfnamefont {C.~N.}\ \bibnamefont {Thomas}}, \bibinfo {author}
  {\bibfnamefont {S.}~\bibnamefont {Withington}}, \bibinfo {author}
  {\bibfnamefont {I.}~\bibnamefont {Siddiqi}}, \bibinfo {author} {\bibfnamefont
  {K.}~\bibnamefont {M{\o}lmer}}, \ and\ \bibinfo {author} {\bibfnamefont
  {J.~J.~L.}\ \bibnamefont {Morton}},\ }\href {\doibase
  10.1103/PhysRevX.12.041014} {\bibfield  {journal} {\bibinfo  {journal}
  {Physical Review X}\ }\textbf {\bibinfo {volume} {12}},\ \bibinfo {pages}
  {041014} (\bibinfo {year} {2022})}\BibitemShut {NoStop}%
\bibitem [{\citenamefont {Imamoglu}(2009)}]{imamoglu_cavity_2009}%
  \BibitemOpen
  \bibfield  {author} {\bibinfo {author} {\bibfnamefont {A.}~\bibnamefont
  {Imamoglu}},\ }\href {\doibase 10.1103/PhysRevLett.102.083602} {\bibfield
  {journal} {\bibinfo  {journal} {Physical Review Letters}\ }\textbf {\bibinfo
  {volume} {102}},\ \bibinfo {pages} {083602} (\bibinfo {year}
  {2009})}\BibitemShut {NoStop}%
\bibitem [{\citenamefont {Kubo}\ \emph {et~al.}(2010)\citenamefont {Kubo},
  \citenamefont {Ong}, \citenamefont {Bertet}, \citenamefont {Vion},
  \citenamefont {Jacques}, \citenamefont {Zheng}, \citenamefont {Dr{\'e}au},
  \citenamefont {Roch}, \citenamefont {Auffeves}, \citenamefont {Jelezko},
  \citenamefont {Wrachtrup}, \citenamefont {Barthe}, \citenamefont {Bergonzo},\
  and\ \citenamefont {Esteve}}]{kubo_strong_2010}%
  \BibitemOpen
  \bibfield  {author} {\bibinfo {author} {\bibfnamefont {Y.}~\bibnamefont
  {Kubo}}, \bibinfo {author} {\bibfnamefont {F.~R.}\ \bibnamefont {Ong}},
  \bibinfo {author} {\bibfnamefont {P.}~\bibnamefont {Bertet}}, \bibinfo
  {author} {\bibfnamefont {D.}~\bibnamefont {Vion}}, \bibinfo {author}
  {\bibfnamefont {V.}~\bibnamefont {Jacques}}, \bibinfo {author} {\bibfnamefont
  {D.}~\bibnamefont {Zheng}}, \bibinfo {author} {\bibfnamefont
  {A.}~\bibnamefont {Dr{\'e}au}}, \bibinfo {author} {\bibfnamefont {J.-F.}\
  \bibnamefont {Roch}}, \bibinfo {author} {\bibfnamefont {A.}~\bibnamefont
  {Auffeves}}, \bibinfo {author} {\bibfnamefont {F.}~\bibnamefont {Jelezko}},
  \bibinfo {author} {\bibfnamefont {J.}~\bibnamefont {Wrachtrup}}, \bibinfo
  {author} {\bibfnamefont {M.~F.}\ \bibnamefont {Barthe}}, \bibinfo {author}
  {\bibfnamefont {P.}~\bibnamefont {Bergonzo}}, \ and\ \bibinfo {author}
  {\bibfnamefont {D.}~\bibnamefont {Esteve}},\ }\href {\doibase
  10.1103/PhysRevLett.105.140502} {\bibfield  {journal} {\bibinfo  {journal}
  {Physical Review Letters}\ }\textbf {\bibinfo {volume} {105}},\ \bibinfo
  {pages} {140502} (\bibinfo {year} {2010})}\BibitemShut {NoStop}%
\bibitem [{\citenamefont {Schuster}\ \emph {et~al.}(2010)\citenamefont
  {Schuster}, \citenamefont {Sears}, \citenamefont {Ginossar}, \citenamefont
  {DiCarlo}, \citenamefont {Frunzio}, \citenamefont {Morton}, \citenamefont
  {Wu}, \citenamefont {Briggs}, \citenamefont {Buckley}, \citenamefont
  {Awschalom},\ and\ \citenamefont
  {Schoelkopf}}]{schuster_high-cooperativity_2010}%
  \BibitemOpen
  \bibfield  {author} {\bibinfo {author} {\bibfnamefont {D.~I.}\ \bibnamefont
  {Schuster}}, \bibinfo {author} {\bibfnamefont {A.~P.}\ \bibnamefont {Sears}},
  \bibinfo {author} {\bibfnamefont {E.}~\bibnamefont {Ginossar}}, \bibinfo
  {author} {\bibfnamefont {L.}~\bibnamefont {DiCarlo}}, \bibinfo {author}
  {\bibfnamefont {L.}~\bibnamefont {Frunzio}}, \bibinfo {author} {\bibfnamefont
  {J.~J.~L.}\ \bibnamefont {Morton}}, \bibinfo {author} {\bibfnamefont
  {H.}~\bibnamefont {Wu}}, \bibinfo {author} {\bibfnamefont {G.~A.~D.}\
  \bibnamefont {Briggs}}, \bibinfo {author} {\bibfnamefont {B.~B.}\
  \bibnamefont {Buckley}}, \bibinfo {author} {\bibfnamefont {D.~D.}\
  \bibnamefont {Awschalom}}, \ and\ \bibinfo {author} {\bibfnamefont {R.~J.}\
  \bibnamefont {Schoelkopf}},\ }\href {\doibase 10.1103/PhysRevLett.105.140501}
  {\bibfield  {journal} {\bibinfo  {journal} {Physical Review Letters}\
  }\textbf {\bibinfo {volume} {105}},\ \bibinfo {pages} {140501} (\bibinfo
  {year} {2010})}\BibitemShut {NoStop}%
\bibitem [{\citenamefont {Abe}\ \emph {et~al.}(2011)\citenamefont {Abe},
  \citenamefont {Wu}, \citenamefont {Ardavan},\ and\ \citenamefont
  {Morton}}]{abe_electron_2011}%
  \BibitemOpen
  \bibfield  {author} {\bibinfo {author} {\bibfnamefont {E.}~\bibnamefont
  {Abe}}, \bibinfo {author} {\bibfnamefont {H.}~\bibnamefont {Wu}}, \bibinfo
  {author} {\bibfnamefont {A.}~\bibnamefont {Ardavan}}, \ and\ \bibinfo
  {author} {\bibfnamefont {J.~J.~L.}\ \bibnamefont {Morton}},\ }\href {\doibase
  10.1063/1.3601930} {\bibfield  {journal} {\bibinfo  {journal} {Applied
  Physics Letters}\ }\textbf {\bibinfo {volume} {98}},\ \bibinfo {pages}
  {251108} (\bibinfo {year} {2011})}\BibitemShut {NoStop}%
\bibitem [{\citenamefont {Zhu}\ \emph {et~al.}(2011)\citenamefont {Zhu},
  \citenamefont {Saito}, \citenamefont {Kemp}, \citenamefont {Kakuyanagi},
  \citenamefont {Karimoto}, \citenamefont {Nakano}, \citenamefont {Munro},
  \citenamefont {Tokura}, \citenamefont {Everitt}, \citenamefont {Nemoto},
  \citenamefont {Kasu}, \citenamefont {Mizuochi},\ and\ \citenamefont
  {Semba}}]{zhu_coherent_2011}%
  \BibitemOpen
  \bibfield  {author} {\bibinfo {author} {\bibfnamefont {X.}~\bibnamefont
  {Zhu}}, \bibinfo {author} {\bibfnamefont {S.}~\bibnamefont {Saito}}, \bibinfo
  {author} {\bibfnamefont {A.}~\bibnamefont {Kemp}}, \bibinfo {author}
  {\bibfnamefont {K.}~\bibnamefont {Kakuyanagi}}, \bibinfo {author}
  {\bibfnamefont {S.-i.}\ \bibnamefont {Karimoto}}, \bibinfo {author}
  {\bibfnamefont {H.}~\bibnamefont {Nakano}}, \bibinfo {author} {\bibfnamefont
  {W.~J.}\ \bibnamefont {Munro}}, \bibinfo {author} {\bibfnamefont
  {Y.}~\bibnamefont {Tokura}}, \bibinfo {author} {\bibfnamefont {M.~S.}\
  \bibnamefont {Everitt}}, \bibinfo {author} {\bibfnamefont {K.}~\bibnamefont
  {Nemoto}}, \bibinfo {author} {\bibfnamefont {M.}~\bibnamefont {Kasu}},
  \bibinfo {author} {\bibfnamefont {N.}~\bibnamefont {Mizuochi}}, \ and\
  \bibinfo {author} {\bibfnamefont {K.}~\bibnamefont {Semba}},\ }\href
  {\doibase 10.1038/nature10462} {\bibfield  {journal} {\bibinfo  {journal}
  {Nature}\ }\textbf {\bibinfo {volume} {478}},\ \bibinfo {pages} {221}
  (\bibinfo {year} {2011})}\BibitemShut {NoStop}%
\bibitem [{\citenamefont {Ams{\"u}ss}\ \emph {et~al.}(2011)\citenamefont
  {Ams{\"u}ss}, \citenamefont {Koller}, \citenamefont {N{\"o}bauer},
  \citenamefont {Putz}, \citenamefont {Rotter}, \citenamefont {Sandner},
  \citenamefont {Schneider}, \citenamefont {Schramb{\"o}ck}, \citenamefont
  {Steinhauser}, \citenamefont {Ritsch}, \citenamefont {Schmiedmayer},\ and\
  \citenamefont {Majer}}]{amsuss_cavity_2011}%
  \BibitemOpen
  \bibfield  {author} {\bibinfo {author} {\bibfnamefont {R.}~\bibnamefont
  {Ams{\"u}ss}}, \bibinfo {author} {\bibfnamefont {C.}~\bibnamefont {Koller}},
  \bibinfo {author} {\bibfnamefont {T.}~\bibnamefont {N{\"o}bauer}}, \bibinfo
  {author} {\bibfnamefont {S.}~\bibnamefont {Putz}}, \bibinfo {author}
  {\bibfnamefont {S.}~\bibnamefont {Rotter}}, \bibinfo {author} {\bibfnamefont
  {K.}~\bibnamefont {Sandner}}, \bibinfo {author} {\bibfnamefont
  {S.}~\bibnamefont {Schneider}}, \bibinfo {author} {\bibfnamefont
  {M.}~\bibnamefont {Schramb{\"o}ck}}, \bibinfo {author} {\bibfnamefont
  {G.}~\bibnamefont {Steinhauser}}, \bibinfo {author} {\bibfnamefont
  {H.}~\bibnamefont {Ritsch}}, \bibinfo {author} {\bibfnamefont
  {J.}~\bibnamefont {Schmiedmayer}}, \ and\ \bibinfo {author} {\bibfnamefont
  {J.}~\bibnamefont {Majer}},\ }\href {\doibase 10.1103/PhysRevLett.107.060502}
  {\bibfield  {journal} {\bibinfo  {journal} {Physical Review Letters}\
  }\textbf {\bibinfo {volume} {107}},\ \bibinfo {pages} {060502} (\bibinfo
  {year} {2011})}\BibitemShut {NoStop}%
\bibitem [{\citenamefont {Ranjan}\ \emph {et~al.}(2013)\citenamefont {Ranjan},
  \citenamefont {de~Lange}, \citenamefont {Schutjens}, \citenamefont
  {Debelhoir}, \citenamefont {Groen}, \citenamefont {Szombati}, \citenamefont
  {Thoen}, \citenamefont {Klapwijk}, \citenamefont {Hanson},\ and\
  \citenamefont {DiCarlo}}]{ranjan_probing_2013}%
  \BibitemOpen
  \bibfield  {author} {\bibinfo {author} {\bibfnamefont {V.}~\bibnamefont
  {Ranjan}}, \bibinfo {author} {\bibfnamefont {G.}~\bibnamefont {de~Lange}},
  \bibinfo {author} {\bibfnamefont {R.}~\bibnamefont {Schutjens}}, \bibinfo
  {author} {\bibfnamefont {T.}~\bibnamefont {Debelhoir}}, \bibinfo {author}
  {\bibfnamefont {J.~P.}\ \bibnamefont {Groen}}, \bibinfo {author}
  {\bibfnamefont {D.}~\bibnamefont {Szombati}}, \bibinfo {author}
  {\bibfnamefont {D.~J.}\ \bibnamefont {Thoen}}, \bibinfo {author}
  {\bibfnamefont {T.~M.}\ \bibnamefont {Klapwijk}}, \bibinfo {author}
  {\bibfnamefont {R.}~\bibnamefont {Hanson}}, \ and\ \bibinfo {author}
  {\bibfnamefont {L.}~\bibnamefont {DiCarlo}},\ }\href {\doibase
  10.1103/PhysRevLett.110.067004} {\bibfield  {journal} {\bibinfo  {journal}
  {Physical Review Letters}\ }\textbf {\bibinfo {volume} {110}},\ \bibinfo
  {pages} {067004} (\bibinfo {year} {2013})}\BibitemShut {NoStop}%
\bibitem [{\citenamefont {Probst}\ \emph {et~al.}(2013)\citenamefont {Probst},
  \citenamefont {Rotzinger}, \citenamefont {W{\"u}nsch}, \citenamefont {Jung},
  \citenamefont {Jerger}, \citenamefont {Siegel}, \citenamefont {Ustinov},\
  and\ \citenamefont {Bushev}}]{probst_anisotropic_2013}%
  \BibitemOpen
  \bibfield  {author} {\bibinfo {author} {\bibfnamefont {S.}~\bibnamefont
  {Probst}}, \bibinfo {author} {\bibfnamefont {H.}~\bibnamefont {Rotzinger}},
  \bibinfo {author} {\bibfnamefont {S.}~\bibnamefont {W{\"u}nsch}}, \bibinfo
  {author} {\bibfnamefont {P.}~\bibnamefont {Jung}}, \bibinfo {author}
  {\bibfnamefont {M.}~\bibnamefont {Jerger}}, \bibinfo {author} {\bibfnamefont
  {M.}~\bibnamefont {Siegel}}, \bibinfo {author} {\bibfnamefont {A.~V.}\
  \bibnamefont {Ustinov}}, \ and\ \bibinfo {author} {\bibfnamefont {P.~A.}\
  \bibnamefont {Bushev}},\ }\href {\doibase 10.1103/PhysRevLett.110.157001}
  {\bibfield  {journal} {\bibinfo  {journal} {Physical Review Letters}\
  }\textbf {\bibinfo {volume} {110}},\ \bibinfo {pages} {157001} (\bibinfo
  {year} {2013})}\BibitemShut {NoStop}%
\bibitem [{\citenamefont {Huebl}\ \emph {et~al.}(2013)\citenamefont {Huebl},
  \citenamefont {Zollitsch}, \citenamefont {Lotze}, \citenamefont {Hocke},
  \citenamefont {Greifenstein}, \citenamefont {Marx}, \citenamefont {Gross},\
  and\ \citenamefont {Goennenwein}}]{huebl_high_2013}%
  \BibitemOpen
  \bibfield  {author} {\bibinfo {author} {\bibfnamefont {H.}~\bibnamefont
  {Huebl}}, \bibinfo {author} {\bibfnamefont {C.~W.}\ \bibnamefont
  {Zollitsch}}, \bibinfo {author} {\bibfnamefont {J.}~\bibnamefont {Lotze}},
  \bibinfo {author} {\bibfnamefont {F.}~\bibnamefont {Hocke}}, \bibinfo
  {author} {\bibfnamefont {M.}~\bibnamefont {Greifenstein}}, \bibinfo {author}
  {\bibfnamefont {A.}~\bibnamefont {Marx}}, \bibinfo {author} {\bibfnamefont
  {R.}~\bibnamefont {Gross}}, \ and\ \bibinfo {author} {\bibfnamefont
  {S.~T.~B.}\ \bibnamefont {Goennenwein}},\ }\href {\doibase
  10.1103/PhysRevLett.111.127003} {\bibfield  {journal} {\bibinfo  {journal}
  {Physical Review Letters}\ }\textbf {\bibinfo {volume} {111}},\ \bibinfo
  {pages} {127003} (\bibinfo {year} {2013})}\BibitemShut {NoStop}%
\bibitem [{\citenamefont {Sigillito}\ \emph {et~al.}(2014)\citenamefont
  {Sigillito}, \citenamefont {Malissa}, \citenamefont {Tyryshkin},
  \citenamefont {Riemann}, \citenamefont {Abrosimov}, \citenamefont {Becker},
  \citenamefont {Pohl}, \citenamefont {Thewalt}, \citenamefont {Itoh},
  \citenamefont {Morton}, \citenamefont {Houck}, \citenamefont {Schuster},\
  and\ \citenamefont {Lyon}}]{sigillito_fast_2014}%
  \BibitemOpen
  \bibfield  {author} {\bibinfo {author} {\bibfnamefont {A.~J.}\ \bibnamefont
  {Sigillito}}, \bibinfo {author} {\bibfnamefont {H.}~\bibnamefont {Malissa}},
  \bibinfo {author} {\bibfnamefont {A.~M.}\ \bibnamefont {Tyryshkin}}, \bibinfo
  {author} {\bibfnamefont {H.}~\bibnamefont {Riemann}}, \bibinfo {author}
  {\bibfnamefont {N.~V.}\ \bibnamefont {Abrosimov}}, \bibinfo {author}
  {\bibfnamefont {P.}~\bibnamefont {Becker}}, \bibinfo {author} {\bibfnamefont
  {H.-J.}\ \bibnamefont {Pohl}}, \bibinfo {author} {\bibfnamefont {M.~L.~W.}\
  \bibnamefont {Thewalt}}, \bibinfo {author} {\bibfnamefont {K.~M.}\
  \bibnamefont {Itoh}}, \bibinfo {author} {\bibfnamefont {J.~J.~L.}\
  \bibnamefont {Morton}}, \bibinfo {author} {\bibfnamefont {A.~A.}\
  \bibnamefont {Houck}}, \bibinfo {author} {\bibfnamefont {D.~I.}\ \bibnamefont
  {Schuster}}, \ and\ \bibinfo {author} {\bibfnamefont {S.~A.}\ \bibnamefont
  {Lyon}},\ }\href {\doibase 10.1063/1.4881613} {\bibfield  {journal} {\bibinfo
   {journal} {Applied Physics Letters}\ }\textbf {\bibinfo {volume} {104}},\
  (\bibinfo {year} {2014})}\BibitemShut {NoStop}%
\bibitem [{\citenamefont {Rose}\ \emph {et~al.}(2017)\citenamefont {Rose},
  \citenamefont {Tyryshkin}, \citenamefont {Riemann}, \citenamefont
  {Abrosimov}, \citenamefont {Becker}, \citenamefont {Pohl}, \citenamefont
  {Thewalt}, \citenamefont {Itoh},\ and\ \citenamefont
  {Lyon}}]{rose_coherent_2017}%
  \BibitemOpen
  \bibfield  {author} {\bibinfo {author} {\bibfnamefont {B.~C.}\ \bibnamefont
  {Rose}}, \bibinfo {author} {\bibfnamefont {A.~M.}\ \bibnamefont {Tyryshkin}},
  \bibinfo {author} {\bibfnamefont {H.}~\bibnamefont {Riemann}}, \bibinfo
  {author} {\bibfnamefont {N.~V.}\ \bibnamefont {Abrosimov}}, \bibinfo {author}
  {\bibfnamefont {P.}~\bibnamefont {Becker}}, \bibinfo {author} {\bibfnamefont
  {H.-J.}\ \bibnamefont {Pohl}}, \bibinfo {author} {\bibfnamefont {M.~L.~W.}\
  \bibnamefont {Thewalt}}, \bibinfo {author} {\bibfnamefont {K.~M.}\
  \bibnamefont {Itoh}}, \ and\ \bibinfo {author} {\bibfnamefont {S.~A.}\
  \bibnamefont {Lyon}},\ }\href {\doibase 10.1103/PhysRevX.7.031002} {\bibfield
   {journal} {\bibinfo  {journal} {Physical Review X}\ }\textbf {\bibinfo
  {volume} {7}},\ \bibinfo {pages} {031002} (\bibinfo {year}
  {2017})}\BibitemShut {NoStop}%
\bibitem [{\citenamefont {Ball}\ \emph {et~al.}(2018)\citenamefont {Ball},
  \citenamefont {Yamashiro}, \citenamefont {Sumiya}, \citenamefont {Onoda},
  \citenamefont {Ohshima}, \citenamefont {Isoya}, \citenamefont
  {Konstantinov},\ and\ \citenamefont {Kubo}}]{ball_loop-gap_2018}%
  \BibitemOpen
  \bibfield  {author} {\bibinfo {author} {\bibfnamefont {J.~R.}\ \bibnamefont
  {Ball}}, \bibinfo {author} {\bibfnamefont {Y.}~\bibnamefont {Yamashiro}},
  \bibinfo {author} {\bibfnamefont {H.}~\bibnamefont {Sumiya}}, \bibinfo
  {author} {\bibfnamefont {S.}~\bibnamefont {Onoda}}, \bibinfo {author}
  {\bibfnamefont {T.}~\bibnamefont {Ohshima}}, \bibinfo {author} {\bibfnamefont
  {J.}~\bibnamefont {Isoya}}, \bibinfo {author} {\bibfnamefont
  {D.}~\bibnamefont {Konstantinov}}, \ and\ \bibinfo {author} {\bibfnamefont
  {Y.}~\bibnamefont {Kubo}},\ }\href {\doibase 10.1063/1.5025744} {\bibfield
  {journal} {\bibinfo  {journal} {Applied Physics Letters}\ }\textbf {\bibinfo
  {volume} {112}},\ \bibinfo {pages} {204102} (\bibinfo {year}
  {2018})}\BibitemShut {NoStop}%
\bibitem [{\citenamefont {Dicke}(1954)}]{dicke_coherence_1954}%
  \BibitemOpen
  \bibfield  {author} {\bibinfo {author} {\bibfnamefont {R.~H.}\ \bibnamefont
  {Dicke}},\ }\href {\doibase 10.1103/PhysRev.93.99} {\bibfield  {journal}
  {\bibinfo  {journal} {Physical Review}\ }\textbf {\bibinfo {volume} {93}},\
  \bibinfo {pages} {99} (\bibinfo {year} {1954})}\BibitemShut {NoStop}%
\bibitem [{\citenamefont {Bloembergen}\ and\ \citenamefont
  {Pound}(1954)}]{bloembergen_radiation_1954}%
  \BibitemOpen
  \bibfield  {author} {\bibinfo {author} {\bibfnamefont {N.}~\bibnamefont
  {Bloembergen}}\ and\ \bibinfo {author} {\bibfnamefont {R.~V.}\ \bibnamefont
  {Pound}},\ }\href {\doibase 10.1103/PhysRev.95.8} {\bibfield  {journal}
  {\bibinfo  {journal} {Physical Review}\ }\textbf {\bibinfo {volume} {95}},\
  \bibinfo {pages} {8} (\bibinfo {year} {1954})}\BibitemShut {NoStop}%
\bibitem [{\citenamefont {Afzelius}\ \emph {et~al.}(2013)\citenamefont
  {Afzelius}, \citenamefont {Sangouard}, \citenamefont {Johansson},
  \citenamefont {Staudt},\ and\ \citenamefont
  {Wilson}}]{afzelius_proposal_2013}%
  \BibitemOpen
  \bibfield  {author} {\bibinfo {author} {\bibfnamefont {M.}~\bibnamefont
  {Afzelius}}, \bibinfo {author} {\bibfnamefont {N.}~\bibnamefont {Sangouard}},
  \bibinfo {author} {\bibfnamefont {G.}~\bibnamefont {Johansson}}, \bibinfo
  {author} {\bibfnamefont {M.~U.}\ \bibnamefont {Staudt}}, \ and\ \bibinfo
  {author} {\bibfnamefont {C.~M.}\ \bibnamefont {Wilson}},\ }\href {\doibase
  10.1088/1367-2630/15/6/065008} {\bibfield  {journal} {\bibinfo  {journal}
  {New Journal of Physics}\ }\textbf {\bibinfo {volume} {15}},\ \bibinfo
  {pages} {065008} (\bibinfo {year} {2013})}\BibitemShut {NoStop}%
\bibitem [{\citenamefont {Mahashabde}\ \emph {et~al.}(2020)\citenamefont
  {Mahashabde}, \citenamefont {Otto}, \citenamefont {Montemurro}, \citenamefont
  {de~Graaf}, \citenamefont {Kubatkin},\ and\ \citenamefont
  {Danilov}}]{mahashabde_fast_2020}%
  \BibitemOpen
  \bibfield  {author} {\bibinfo {author} {\bibfnamefont {S.}~\bibnamefont
  {Mahashabde}}, \bibinfo {author} {\bibfnamefont {E.}~\bibnamefont {Otto}},
  \bibinfo {author} {\bibfnamefont {D.}~\bibnamefont {Montemurro}}, \bibinfo
  {author} {\bibfnamefont {S.}~\bibnamefont {de~Graaf}}, \bibinfo {author}
  {\bibfnamefont {S.}~\bibnamefont {Kubatkin}}, \ and\ \bibinfo {author}
  {\bibfnamefont {A.}~\bibnamefont {Danilov}},\ }\href {\doibase
  10.1103/PhysRevApplied.14.044040} {\bibfield  {journal} {\bibinfo  {journal}
  {Physical Review Applied}\ }\textbf {\bibinfo {volume} {14}},\ \bibinfo
  {pages} {044040} (\bibinfo {year} {2020})}\BibitemShut {NoStop}%
\bibitem [{\citenamefont {Ranjan}\ \emph {et~al.}(2022)\citenamefont {Ranjan},
  \citenamefont {Wen}, \citenamefont {Keyser}, \citenamefont {Kubatkin},
  \citenamefont {Danilov}, \citenamefont {Lindstr{\"o}m}, \citenamefont
  {Bertet},\ and\ \citenamefont {de~Graaf}}]{ranjan_spin-echo_2022}%
  \BibitemOpen
  \bibfield  {author} {\bibinfo {author} {\bibfnamefont {V.}~\bibnamefont
  {Ranjan}}, \bibinfo {author} {\bibfnamefont {Y.}~\bibnamefont {Wen}},
  \bibinfo {author} {\bibfnamefont {A.~K.~V.}\ \bibnamefont {Keyser}}, \bibinfo
  {author} {\bibfnamefont {S.~E.}\ \bibnamefont {Kubatkin}}, \bibinfo {author}
  {\bibfnamefont {A.~V.}\ \bibnamefont {Danilov}}, \bibinfo {author}
  {\bibfnamefont {T.}~\bibnamefont {Lindstr{\"o}m}}, \bibinfo {author}
  {\bibfnamefont {P.}~\bibnamefont {Bertet}}, \ and\ \bibinfo {author}
  {\bibfnamefont {S.~E.}\ \bibnamefont {de~Graaf}},\ }\href {\doibase
  10.1103/PhysRevLett.129.180504} {\bibfield  {journal} {\bibinfo  {journal}
  {Physical Review Letters}\ }\textbf {\bibinfo {volume} {129}},\ \bibinfo
  {pages} {180504} (\bibinfo {year} {2022})}\BibitemShut {NoStop}%
\bibitem [{\citenamefont {Rakonjac}\ \emph {et~al.}(2020)\citenamefont
  {Rakonjac}, \citenamefont {Chen}, \citenamefont {Horvath},\ and\
  \citenamefont {Longdell}}]{rakonjac_long_2020}%
  \BibitemOpen
  \bibfield  {author} {\bibinfo {author} {\bibfnamefont {J.~V.}\ \bibnamefont
  {Rakonjac}}, \bibinfo {author} {\bibfnamefont {Y.-H.}\ \bibnamefont {Chen}},
  \bibinfo {author} {\bibfnamefont {S.~P.}\ \bibnamefont {Horvath}}, \ and\
  \bibinfo {author} {\bibfnamefont {J.~J.}\ \bibnamefont {Longdell}},\ }\href
  {\doibase 10.1103/PhysRevB.101.184430} {\bibfield  {journal} {\bibinfo
  {journal} {Physical Review B}\ }\textbf {\bibinfo {volume} {101}},\ \bibinfo
  {pages} {184430} (\bibinfo {year} {2020})}\BibitemShut {NoStop}%
\bibitem [{\citenamefont {Enrique}(1971)}]{enrique_optical_1971}%
  \BibitemOpen
  \bibfield  {author} {\bibinfo {author} {\bibfnamefont {B.~G.}\ \bibnamefont
  {Enrique}},\ }\href {\doibase 10.1063/1.1676445} {\bibfield  {journal}
  {\bibinfo  {journal} {The Journal of Chemical Physics}\ }\textbf {\bibinfo
  {volume} {55}},\ \bibinfo {pages} {2538} (\bibinfo {year}
  {1971})}\BibitemShut {NoStop}%
\bibitem [{\citenamefont {Mims}\ and\ \citenamefont
  {Gillen}(1966)}]{mims_broadening_1966}%
  \BibitemOpen
  \bibfield  {author} {\bibinfo {author} {\bibfnamefont {W.~B.}\ \bibnamefont
  {Mims}}\ and\ \bibinfo {author} {\bibfnamefont {R.}~\bibnamefont {Gillen}},\
  }\href {\doibase 10.1103/PhysRev.148.438} {\bibfield  {journal} {\bibinfo
  {journal} {Physical Review}\ }\textbf {\bibinfo {volume} {148}},\ \bibinfo
  {pages} {438} (\bibinfo {year} {1966})}\BibitemShut {NoStop}%
\bibitem [{\citenamefont {Dantec}\ \emph {et~al.}(2021)\citenamefont {Dantec},
  \citenamefont {Ran{\v c}i{\'c}}, \citenamefont {Lin}, \citenamefont
  {Billaud}, \citenamefont {Ranjan}, \citenamefont {Flanigan}, \citenamefont
  {Bertaina}, \citenamefont {Chaneli{\`e}re}, \citenamefont {Goldner},
  \citenamefont {Erb}, \citenamefont {Liu}, \citenamefont {Est{\`e}ve},
  \citenamefont {Vion}, \citenamefont {Flurin},\ and\ \citenamefont
  {Bertet}}]{dantec_twenty-threemillisecond_2021}%
  \BibitemOpen
  \bibfield  {author} {\bibinfo {author} {\bibfnamefont {M.~L.}\ \bibnamefont
  {Dantec}}, \bibinfo {author} {\bibfnamefont {M.}~\bibnamefont {Ran{\v
  c}i{\'c}}}, \bibinfo {author} {\bibfnamefont {S.}~\bibnamefont {Lin}},
  \bibinfo {author} {\bibfnamefont {E.}~\bibnamefont {Billaud}}, \bibinfo
  {author} {\bibfnamefont {V.}~\bibnamefont {Ranjan}}, \bibinfo {author}
  {\bibfnamefont {D.}~\bibnamefont {Flanigan}}, \bibinfo {author}
  {\bibfnamefont {S.}~\bibnamefont {Bertaina}}, \bibinfo {author}
  {\bibfnamefont {T.}~\bibnamefont {Chaneli{\`e}re}}, \bibinfo {author}
  {\bibfnamefont {P.}~\bibnamefont {Goldner}}, \bibinfo {author} {\bibfnamefont
  {A.}~\bibnamefont {Erb}}, \bibinfo {author} {\bibfnamefont {R.~B.}\
  \bibnamefont {Liu}}, \bibinfo {author} {\bibfnamefont {D.}~\bibnamefont
  {Est{\`e}ve}}, \bibinfo {author} {\bibfnamefont {D.}~\bibnamefont {Vion}},
  \bibinfo {author} {\bibfnamefont {E.}~\bibnamefont {Flurin}}, \ and\ \bibinfo
  {author} {\bibfnamefont {P.}~\bibnamefont {Bertet}},\ }\href {\doibase
  10.1126/sciadv.abj9786} {\bibfield  {journal} {\bibinfo  {journal} {Science
  Advances}\ } (\bibinfo {year} {2021}),\ 10.1126/sciadv.abj9786}\BibitemShut
  {NoStop}%
\bibitem [{\citenamefont {Ranjan}\ \emph
  {et~al.}(2020{\natexlab{c}})\citenamefont {Ranjan}, \citenamefont {Probst},
  \citenamefont {Albanese}, \citenamefont {Doll}, \citenamefont {Jacquot},
  \citenamefont {Flurin}, \citenamefont {Heeres}, \citenamefont {Vion},
  \citenamefont {Esteve}, \citenamefont {Morton},\ and\ \citenamefont
  {Bertet}}]{ranjan_pulsed_2020}%
  \BibitemOpen
  \bibfield  {author} {\bibinfo {author} {\bibfnamefont {V.}~\bibnamefont
  {Ranjan}}, \bibinfo {author} {\bibfnamefont {S.}~\bibnamefont {Probst}},
  \bibinfo {author} {\bibfnamefont {B.}~\bibnamefont {Albanese}}, \bibinfo
  {author} {\bibfnamefont {A.}~\bibnamefont {Doll}}, \bibinfo {author}
  {\bibfnamefont {O.}~\bibnamefont {Jacquot}}, \bibinfo {author} {\bibfnamefont
  {E.}~\bibnamefont {Flurin}}, \bibinfo {author} {\bibfnamefont
  {R.}~\bibnamefont {Heeres}}, \bibinfo {author} {\bibfnamefont
  {D.}~\bibnamefont {Vion}}, \bibinfo {author} {\bibfnamefont {D.}~\bibnamefont
  {Esteve}}, \bibinfo {author} {\bibfnamefont {J.~J.~L.}\ \bibnamefont
  {Morton}}, \ and\ \bibinfo {author} {\bibfnamefont {P.}~\bibnamefont
  {Bertet}},\ }\href {\doibase https://doi.org/10.1016/j.jmr.2019.106662}
  {\bibfield  {journal} {\bibinfo  {journal} {Journal of Magnetic Resonance}\
  }\textbf {\bibinfo {volume} {310}},\ \bibinfo {pages} {106662} (\bibinfo
  {year} {2020}{\natexlab{c}})}\BibitemShut {NoStop}%
\bibitem [{\citenamefont {O{\textquoteright}Sullivan}\ \emph
  {et~al.}(2020)\citenamefont {O{\textquoteright}Sullivan}, \citenamefont
  {Kennedy}, \citenamefont {Zollitsch}, \citenamefont {{\v S}im{\.e}nas},
  \citenamefont {Thomas}, \citenamefont {Abdurakhimov}, \citenamefont
  {Withington},\ and\ \citenamefont {Morton}}]{osullivan_spin-resonance_2020}%
  \BibitemOpen
  \bibfield  {author} {\bibinfo {author} {\bibfnamefont {J.}~\bibnamefont
  {O{\textquoteright}Sullivan}}, \bibinfo {author} {\bibfnamefont {O.~W.}\
  \bibnamefont {Kennedy}}, \bibinfo {author} {\bibfnamefont {C.~W.}\
  \bibnamefont {Zollitsch}}, \bibinfo {author} {\bibfnamefont {M.}~\bibnamefont
  {{\v S}im{\.e}nas}}, \bibinfo {author} {\bibfnamefont {C.~N.}\ \bibnamefont
  {Thomas}}, \bibinfo {author} {\bibfnamefont {L.~V.}\ \bibnamefont
  {Abdurakhimov}}, \bibinfo {author} {\bibfnamefont {S.}~\bibnamefont
  {Withington}}, \ and\ \bibinfo {author} {\bibfnamefont {J.~J.}\ \bibnamefont
  {Morton}},\ }\href {\doibase 10.1103/PhysRevApplied.14.064050} {\bibfield
  {journal} {\bibinfo  {journal} {Physical Review Applied}\ }\textbf {\bibinfo
  {volume} {14}},\ \bibinfo {pages} {064050} (\bibinfo {year}
  {2020})}\BibitemShut {NoStop}%
\bibitem [{\citenamefont {Dantec}(2022)}]{dantec_notitle_2022}%
  \BibitemOpen
  \bibfield  {author} {\bibinfo {author} {\bibfnamefont {M.~L.}\ \bibnamefont
  {Dantec}},\ }\href {https://tel.archives-ouvertes.fr/tel-03579857} {\bibinfo
  {type} {phd thesis}},\ \bibinfo  {school} {Universit{\'e} Paris-Saclay}
  (\bibinfo {year} {2022})\BibitemShut {NoStop}%
\bibitem [{\citenamefont {Ran{\v c}i{\'c}}\ \emph {et~al.}(2022)\citenamefont
  {Ran{\v c}i{\'c}}, \citenamefont {Le~Dantec}, \citenamefont {Lin},
  \citenamefont {Bertaina}, \citenamefont {Chaneli{\`e}re}, \citenamefont
  {Serrano}, \citenamefont {Goldner}, \citenamefont {Liu}, \citenamefont
  {Flurin}, \citenamefont {Est{\`e}ve}, \citenamefont {Vion},\ and\
  \citenamefont {Bertet}}]{rancic_electron-spin_2022-1}%
  \BibitemOpen
  \bibfield  {author} {\bibinfo {author} {\bibfnamefont {M.}~\bibnamefont
  {Ran{\v c}i{\'c}}}, \bibinfo {author} {\bibfnamefont {M.}~\bibnamefont
  {Le~Dantec}}, \bibinfo {author} {\bibfnamefont {S.}~\bibnamefont {Lin}},
  \bibinfo {author} {\bibfnamefont {S.}~\bibnamefont {Bertaina}}, \bibinfo
  {author} {\bibfnamefont {T.}~\bibnamefont {Chaneli{\`e}re}}, \bibinfo
  {author} {\bibfnamefont {D.}~\bibnamefont {Serrano}}, \bibinfo {author}
  {\bibfnamefont {P.}~\bibnamefont {Goldner}}, \bibinfo {author} {\bibfnamefont
  {R.~B.}\ \bibnamefont {Liu}}, \bibinfo {author} {\bibfnamefont
  {E.}~\bibnamefont {Flurin}}, \bibinfo {author} {\bibfnamefont
  {D.}~\bibnamefont {Est{\`e}ve}}, \bibinfo {author} {\bibfnamefont
  {D.}~\bibnamefont {Vion}}, \ and\ \bibinfo {author} {\bibfnamefont
  {P.}~\bibnamefont {Bertet}},\ }\href {\doibase 10.1103/PhysRevB.106.144412}
  {\bibfield  {journal} {\bibinfo  {journal} {Physical Review B}\ }\textbf
  {\bibinfo {volume} {106}},\ \bibinfo {pages} {144412} (\bibinfo {year}
  {2022})}\BibitemShut {NoStop}%
\bibitem [{\citenamefont {Alexander}\ \emph {et~al.}(2022)\citenamefont
  {Alexander}, \citenamefont {Dold}, \citenamefont {Kennedy}, \citenamefont
  {{\v S}im{\.e}nas}, \citenamefont {O'Sullivan}, \citenamefont {Zollitsch},
  \citenamefont {Welinski}, \citenamefont {Ferrier}, \citenamefont
  {Lafitte-Houssat}, \citenamefont {Lindstr{\"o}m}, \citenamefont {Goldner},\
  and\ \citenamefont {Morton}}]{alexander_coherent_2022}%
  \BibitemOpen
  \bibfield  {author} {\bibinfo {author} {\bibfnamefont {J.}~\bibnamefont
  {Alexander}}, \bibinfo {author} {\bibfnamefont {G.}~\bibnamefont {Dold}},
  \bibinfo {author} {\bibfnamefont {O.~W.}\ \bibnamefont {Kennedy}}, \bibinfo
  {author} {\bibfnamefont {M.}~\bibnamefont {{\v S}im{\.e}nas}}, \bibinfo
  {author} {\bibfnamefont {J.}~\bibnamefont {O'Sullivan}}, \bibinfo {author}
  {\bibfnamefont {C.~W.}\ \bibnamefont {Zollitsch}}, \bibinfo {author}
  {\bibfnamefont {S.}~\bibnamefont {Welinski}}, \bibinfo {author}
  {\bibfnamefont {A.}~\bibnamefont {Ferrier}}, \bibinfo {author} {\bibfnamefont
  {E.}~\bibnamefont {Lafitte-Houssat}}, \bibinfo {author} {\bibfnamefont
  {T.}~\bibnamefont {Lindstr{\"o}m}}, \bibinfo {author} {\bibfnamefont
  {P.}~\bibnamefont {Goldner}}, \ and\ \bibinfo {author} {\bibfnamefont
  {J.~J.~L.}\ \bibnamefont {Morton}},\ }\href {\doibase
  10.1103/PhysRevB.106.245416} {\bibfield  {journal} {\bibinfo  {journal}
  {Physical Review B}\ }\textbf {\bibinfo {volume} {106}},\ \bibinfo {pages}
  {245416} (\bibinfo {year} {2022})}\BibitemShut {NoStop}%
\bibitem [{\citenamefont {Wu}\ \emph {et~al.}(2010)\citenamefont {Wu},
  \citenamefont {George}, \citenamefont {Wesenberg}, \citenamefont {Moelmer},
  \citenamefont {Schuster}, \citenamefont {Schoelkopf}, \citenamefont {Itoh},
  \citenamefont {Ardavan}, \citenamefont {Morton},\ and\ \citenamefont
  {Briggs}}]{wu_storage_2010}%
  \BibitemOpen
  \bibfield  {author} {\bibinfo {author} {\bibfnamefont {H.}~\bibnamefont
  {Wu}}, \bibinfo {author} {\bibfnamefont {R.~E.}\ \bibnamefont {George}},
  \bibinfo {author} {\bibfnamefont {J.~H.}\ \bibnamefont {Wesenberg}}, \bibinfo
  {author} {\bibfnamefont {K.}~\bibnamefont {Moelmer}}, \bibinfo {author}
  {\bibfnamefont {D.~I.}\ \bibnamefont {Schuster}}, \bibinfo {author}
  {\bibfnamefont {R.~J.}\ \bibnamefont {Schoelkopf}}, \bibinfo {author}
  {\bibfnamefont {K.~M.}\ \bibnamefont {Itoh}}, \bibinfo {author}
  {\bibfnamefont {A.}~\bibnamefont {Ardavan}}, \bibinfo {author} {\bibfnamefont
  {J.~J.~L.}\ \bibnamefont {Morton}}, \ and\ \bibinfo {author} {\bibfnamefont
  {G.~A.~D.}\ \bibnamefont {Briggs}},\ }\href {\doibase
  10.1103/PhysRevLett.105.140503} {\bibfield  {journal} {\bibinfo  {journal}
  {Physical Review Letters}\ }\textbf {\bibinfo {volume} {105}},\ \bibinfo
  {pages} {140503} (\bibinfo {year} {2010})}\BibitemShut {NoStop}%
\bibitem [{\citenamefont {Burstein}(1996)}]{burstein_stimulated_1996}%
  \BibitemOpen
  \bibfield  {author} {\bibinfo {author} {\bibfnamefont {D.}~\bibnamefont
  {Burstein}},\ }\href {\doibase
  10.1002/(SICI)1099-0534(1996)8:4<269::AID-CMR3>3.0.CO;2-X} {\bibfield
  {journal} {\bibinfo  {journal} {Concepts in Magnetic Resonance}\ }\textbf
  {\bibinfo {volume} {8}},\ \bibinfo {pages} {269} (\bibinfo {year}
  {1996})}\BibitemShut {NoStop}%
\end{thebibliography}%

\end{document}